\documentclass[aps,pra,reprint,superscriptaddress,nofootinbib]{revtex4-2}

\usepackage{amsmath,amssymb,bm}
\usepackage{xcolor}
\usepackage[normalem]{ulem}
\usepackage{graphicx}
\usepackage{hyperref}

\usepackage{lineno}

\newif\iferbedit
\erbeditfalse

\iferbedit
  
  \newcommand{\erbdel}[1]{\textcolor{red}{\sout{#1}}}
  
\else
  
  \newcommand{\erbdel}[1]{}
  
\fi

\newcommand{\Tr}{\operatorname{Tr}}
\newcommand{\dd}{\mathrm{d}}
\newcommand{\rhoSS}{\rho_{\rm ss}}

\begin{document}
%\linenumbers

%\title{Geometric Origin of Complementarity Constraints in Open Quantum Systems}
\title{Holonomy and Complementarity in Open Quantum Systems}

\author{Eric R. Bittner}
\affiliation{Department of Physics, University of Houston, Houston, Texas 77204, USA}

\date{\today}

\begin{abstract}
Complementarity relations constrain the distribution of coherence, predictability, and openness in quantum systems. Here we show that, in open quantum systems, these local constraints acquire a geometric interpretation through quasistatic transport. For a driven dissipative qubit, the complementarity variables define cylindrical coordinates on the Bloch sphere, while openness appears geometrically as a radial deficit associated with reduction from a larger Hilbert space. Quasistatic driving induces a work connection on the resulting steady-state manifold whose curvature determines the cyclic response. Hamiltonian-aligned dissipation produces an exact work connection and vanishing cyclic work, whereas fixed pointer-basis dissipation generates non-integrable transport, finite curvature, and holonomic response. The resulting curvature admits a phase-resolved representation on the triality manifold and develops perturbatively with pointer--Hamiltonian mismatch. In the weak-mismatch limit, the curvature is governed by a competition between coherence-preserving and pure-dephasing channels, producing symmetry-related positive- and negative-curvature sectors. These results establish a direct connection between complementarity, dissipation, and geometric thermodynamic response, and show that cyclic quasistatic work provides an operational probe of nonequilibrium quantum geometry.
\end{abstract}

\maketitle

\section{Introduction.}

Complementarity relations are traditionally interpreted as
kinematic constraints limiting the simultaneous manifestation
of coherence, $C$, predictability, $P$, and entanglement, $E$. Beginning with wave--particle duality relations and their later generalizations to multipartite and open quantum systems, such constraints quantify the tradeoffs governing the
accessible information content of quantum states
\cite{Greenberger1988duality,Englert1996,Jakob2002,Qian2016coherenceConstraints,Swain:2026aa}.
In open quantum systems, however, coherence and entanglement evolve continuously under the combined action
of coherent driving and dissipation, raising a deeper
question: \emph{are complementarity relations merely preserved by
the dynamics, or do they themselves define an underlying
geometric structure governing nonequilibrium response?}

Previous work has largely treated complementarity relations
as algebraic constraints on admissible quantum states.
Here, we show that the complementarity variables
$(C,P,E)$ admit a direct geometric interpretation.
Specifically, we demonstrate that these variables define
cylindrical coordinates on the Bloch sphere, with
entanglement appearing geometrically as a radial deficit
associated with a reduction from a larger Hilbert space.
This identification elevates complementarity from a purely
kinematic relation to a coordinate system on state space.
Related geometric formulations have also appeared in nonequilibrium thermodynamics and thermodynamic length approaches, where dissipation and response are characterized through geometric structures on control manifolds\cite{Crooks1999,Ruppeiner1995,Harbola2016geometric}.

The central result of this paper is that this local
structure acquires a global physical meaning under
quasistatic transport. Geometric phases, holonomy, and curvature have long played a central role in adiabatic
quantum transport and response theory
\cite{Berry1984,Simon1983,
Provost1980,Wilczek1984,Xiao2010}.
More recently, related geometric structures have emerged in driven open quantum systems and nonequilibrium steady-state
dynamics, where dissipation and basis mismatch generate
nontrivial geometric response
\cite{Sarandy2005,Avron2009,
Rahav2012,Bittner2026_oqs,Bittner2026_jcp,Bittner2026_classical}.
For driven open quantum systems, the steady state defines a
work one-form
\begin{equation}
A_i(\lambda)
=
\mathrm{Tr}\!\left[
\rho_{\rm ss}(\lambda)\,\partial_i H(\lambda)
\right],
\end{equation}
whose integral over a closed cycle yields the quasistatic
work.

The geometric structure developed here differs from conventional quantum geometric constructions based on Berry curvature, the quantum geometric tensor, or information-theoretic metrics on density operators. Those approaches characterize the geometry of quantum states or parameter manifolds directly. By contrast, the present curvature arises from the quasistatic transport of nonequilibrium steady states under dissipative dynamics and is operationally measured through cyclic work. The resulting geometry is therefore not purely kinematic, but intrinsically dynamical and thermodynamic.

When the steady state is Gibbsian, this connection is exact, and the quasistatic work is path-independent. However, when the dissipative pointer basis is not aligned with the
instantaneous Hamiltonian eigenbasis, the connection
becomes non-integrable and develops curvature. In this case, cyclic driving protocols generate finite work, and the quasistatic response becomes a holonomy, \emph{i.e.}, a
path-dependent geometric response generated by cyclic
transport over control space.\cite{Bittner2026_oqs,Bittner2026_jcp}

This observation promotes complementarity from a constraint relation on admissible states to a \emph{geometric
transport structure governing nonequilibrium response.} The triality relation constrains the local coordinates
of the state, while the curvature of the work
connection governs their global transport over a control
manifold. In this sense, complementarity defines the local
coordinate structure of the state manifold, while holonomy
governs the global response generated by transport through
control space.

\begin{figure*}[t]
\centering
\includegraphics[width=0.95\textwidth]{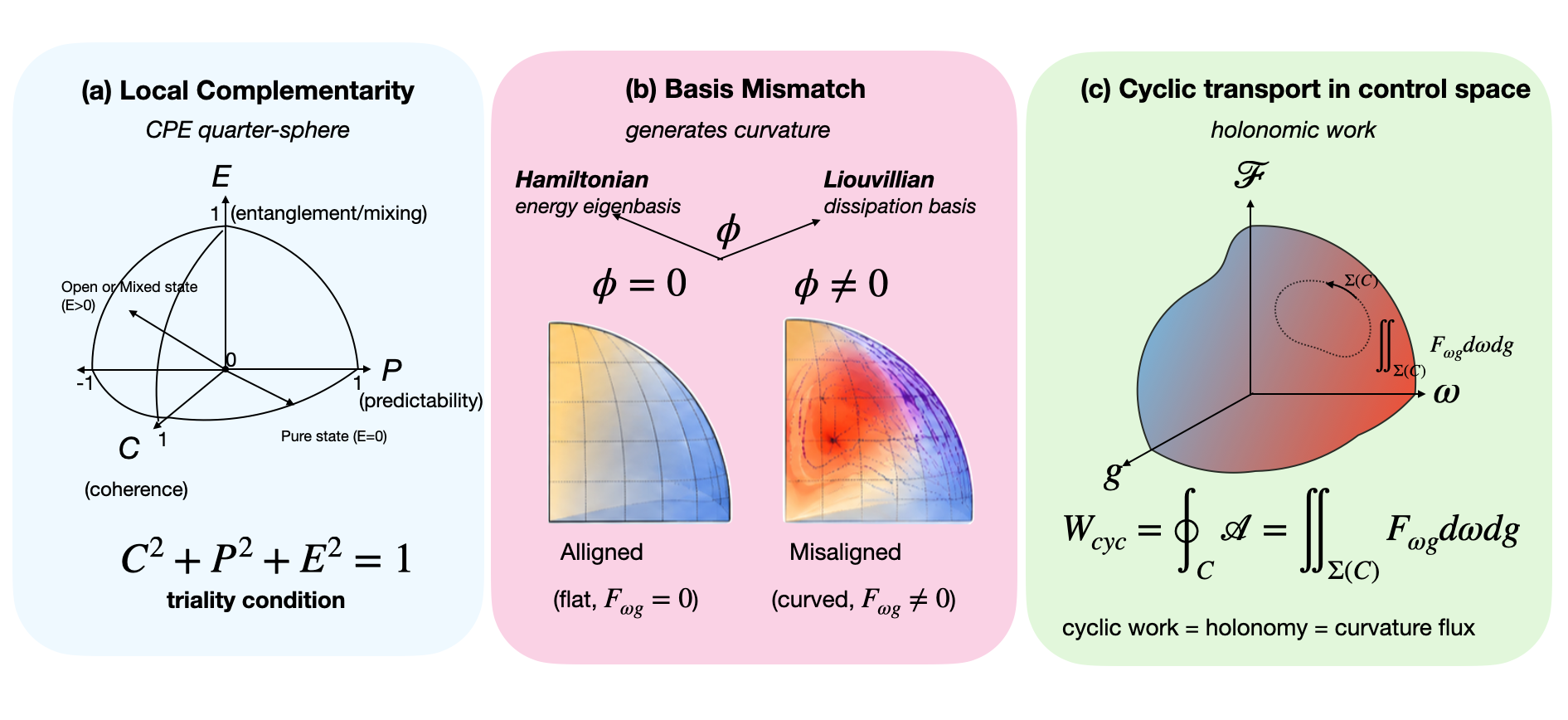}
\caption{
\textbf{Geometric interpretation of complementarity transport in open quantum systems.}
(a) The complementarity variables coherence, predictability, and openness define a constrained triality manifold associated with the reduced quantum state.
(b) Misalignment between the Hamiltonian eigenbasis and the dissipative pointer basis generates curvature on the steady-state manifold. Aligned dynamics produce a locally exact work connection with vanishing curvature, while basis mismatch produces non-integrable transport and finite holonomic response.
(c) Cyclic quasistatic transport in control space accumulates finite work determined by the curvature flux through the enclosed surface, illustrating the geometric origin of nonequilibrium response.
}
\label{fig:triality_concept}
\end{figure*}

We demonstrate these ideas explicitly for a driven
dissipative qubit. We show that Hamiltonian-aligned
dissipation produces an integrable work connection and
vanishing cyclic work, whereas fixed pointer-basis
dissipation produces finite curvature and nontrivial
holonomic response. The resulting curvature can be
expressed as a phase-resolved function of the
complementarity variables, revealing that quasistatic work emerges from the
geometric transport of coherence, predictability, and
entanglement.

These results establish a direct connection between
complementarity and thermodynamic response in open quantum
systems. More broadly, they suggest that information-theoretic constraints are naturally embedded
within a geometric framework in which observable quantities,
such as cyclic work, directly probe the curvature of
state-space transport.

%\paragraph*{Positioning and scope.}

Recent work has emphasized the persistence of
complementarity relations under open-system dynamics,
demonstrating that tradeoff constraints between coherence,
predictability, and entanglement can survive the presence
of noise and dissipation
\cite{Jakob2002,Qian2016coherenceConstraints,Swain:2026aa}.
While these studies establish the robustness of complementarity as a kinematic constraint, they do not
address the dynamical or geometric origin of these
relations.

Here, we take a different perspective
\cite{Bittner2026_oqs,Bittner2026_jcp}.
Rather than treating complementarity as a constraint that
must be preserved under evolution, we show that it arises naturally as the local coordinate structure of a
nonequilibrium steady-state manifold. Within this
framework, transport of the complementarity variables over a control manifold induces a work connection whose
curvature determines the quasistatic response. In this
sense, complementarity is not simply maintained by the dynamics but is embedded within a geometric structure that
directly governs observable quantities.

This distinction is essential. The local triality relation
\[
C^2+P^2+E^2=1
\]
is fundamentally kinematic, defining the admissible local coordinate structure of the state manifold. 
By contrast, pointer-basis mismatch generates curvature on the
steady-state manifold, producing curvature-induced transport and finite holonomic response.  Finite cyclic work
therefore provides a direct operational probe of the
non-integrability of complementarity transport in open
quantum systems.

Figure~\ref{fig:triality_concept} summarizes the geometric structure developed in this work. The complementarity variables coherence $C$, predictability $P$, and openness coordinate $E$ define a constrained manifold associated with the reduced qubit state. Here $C$ measures coherence within the pointer-basis plane, $P$ measures population imbalance, and $E$ quantifies Bloch-sphere contraction induced by openness and mixing. Together, these variables define a restricted quarter-sphere geometry representing the admissible reduced-state manifold.

Within this framework, complementarity acquires a direct geometric interpretation. The local triality relation defines the coordinate structure of the steady-state manifold, while dissipative dynamics determine how those coordinates are transported under quasistatic driving. When the dissipative pointer basis is aligned with the Hamiltonian eigenbasis, the work connection is locally exact and cyclic transport produces no net work. By contrast, pointer-basis mismatch generates curvature on the steady-state manifold, producing non-integrable transport and finite holonomic response.

The final panel illustrates the operational consequence of this geometry. A closed quasistatic cycle in control space accumulates finite work determined by the curvature flux enclosed by the cycle. Cyclic quasistatic work therefore provides a direct probe of the geometric transport of coherence, predictability, and openness in driven open quantum systems.
\begin{figure*}[t]
\centering

\begin{minipage}{0.48\textwidth}
    \centering
    \includegraphics[width=\linewidth]{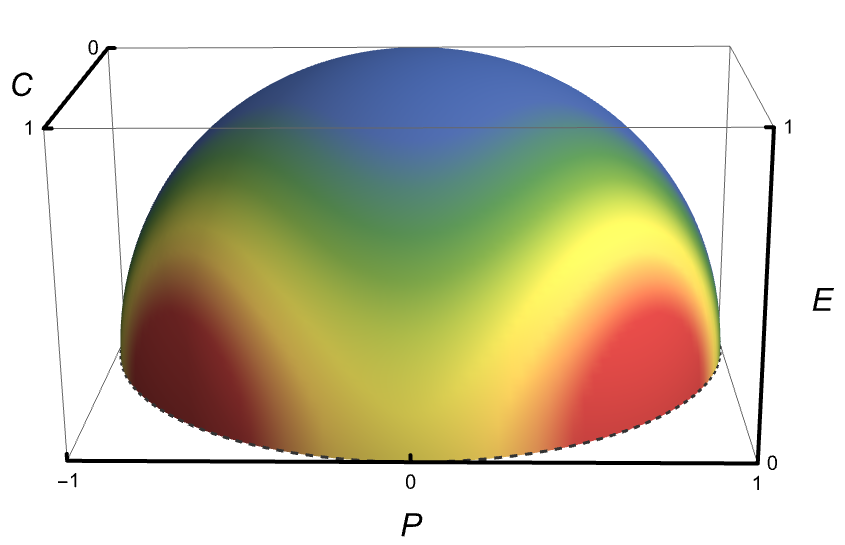}
\end{minipage}
\hfill
\begin{minipage}{0.48\textwidth}
    \centering
    \includegraphics[width=\linewidth]{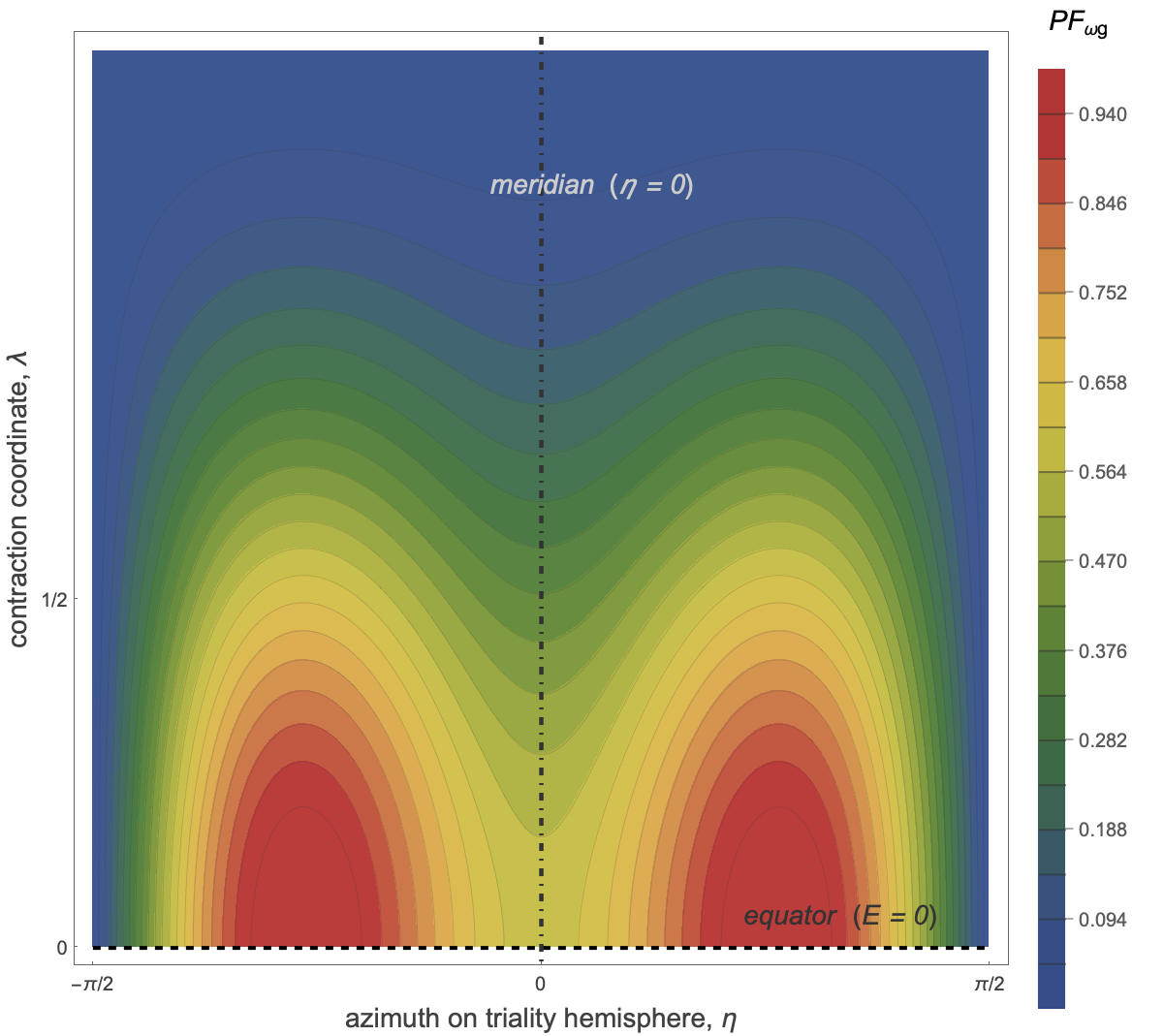}
\end{minipage}

\caption{
(a) Regularized curvature field $P F_{\omega g}$ painted on the physical
triality quarter-sphere defined by
$C^2 + P^2 + E^2 = 1$ with $C \ge 0$ for
$\phi=\pi/2$.  Here $C$ denotes coherence,
$P$ the predictability (population imbalance), and
$E=\sqrt{1-(C^2+P^2)}$ the Bloch-sphere contraction coordinate that
quantifies openness and mixedness.  The dashed meridian marks the
$E=0$ boundary corresponding to pure states without Bloch-sphere
contraction.  The curvature localizes into symmetry-related interior
regions of the manifold, demonstrating that holonomic response is
maximized not at extremal limits, but in partially contracted states
where coherence, predictability, and mixing coexist.
(b) Mercator-like projection of the same curvature field onto the
$(\eta,\lambda)$ coordinate chart of the triality manifold, where
$\eta$ parameterizes azimuthal position around the
coherence--predictability circle and
$E=\sin\lambda$.  The dashed horizontal line denotes the equator
$E=0$, while the vertical meridian $\eta=0$ marks the symmetry axis
separating the two curvature sectors.  The contour topology reveals
two symmetry-related curvature basins connected through a low-curvature
corridor near the pure-state boundary, illustrating the geometric
organization of holonomic work on the complementarity manifold.
}
\label{fig:triality_curvature}
\end{figure*}
%\paragraph*{Model.}

\section{Triality Geometry of Open Quantum States}

We now develop the geometric structure underlying complementarity in driven open quantum systems. The complementarity variables coherence, predictability, and openness define coordinates on a constrained steady-state manifold, while quasistatic transport on this manifold generates geometric response. We illustrate these ideas explicitly for a driven dissipative qubit, where the competition between coherent precession and fixed pointer-basis dissipation produces a nonequilibrium steady state with nontrivial curvature. Although we focus on the minimal two-level setting, the underlying geometric structure is not restricted to single qubits. More generally, complementarity constraints in higher-dimensional and many-body open quantum systems likewise define constrained state-space manifolds whose transport properties generate geometric response under driven dissipative dynamics. The single qubit, therefore, provides the simplest realization of a broader nonequilibrium geometric framework connecting complementarity, dissipation, and holonomic transport.

\subsection{A model qubit}
We consider a driven qubit with Hamiltonian
\begin{equation}
H(\omega,g)
=
\frac{1}{2}
\left(
\omega\sigma_z+g\sigma_x
\right),
\label{eq:H}
\end{equation}
where $\omega$ and $g$ are externally controlled parameters. The reduced qubit state is represented in Bloch form,
\begin{equation}
\rho
=
\frac{1}{2}
\left(
I+x\sigma_x+y\sigma_y+z\sigma_z
\right)
=
\frac{1}{2}
\left(
I+\bm r\cdot \bm \sigma
\right).
\label{eq:bloch}
\end{equation}
For fixed pointer-basis relaxation and dephasing, the Bloch equation is
\begin{equation}
\dot{\bm r}
=
\bm h\times \bm r
-
\Gamma_2
\left(
x\hat{\bm x}+y\hat{\bm y}
\right)
-
\Gamma_1
\left(
z-z_0
\right)
\hat{\bm z},
\label{eq:bloch-eom}
\end{equation}
with
\begin{equation}
\bm h=(g,0,\omega).
\end{equation}

\paragraph{Triality geometry.}
Define cylindrical Bloch coordinates
\begin{equation}
C=\sqrt{x^2+y^2},
\qquad
P=z,
\qquad
R=\sqrt{x^2+y^2+z^2}.
\label{eq:C-P-R}
\end{equation}
The quantity $C$ measures coherence in the pointer-basis plane,
while $P$ measures predictability (population imbalance).
Physical states satisfy $R\leq1$, so that
\begin{equation}
C^2+P^2=R^2\leq1.
\label{eq:duality}
\end{equation}
Introducing the radial deficit
\begin{equation}
E=\sqrt{1-R^2},
\label{eq:E-def}
\end{equation}
which quantifies Bloch-sphere contraction induced by openness and
mixing, the Bloch-sphere constraint acquires the normalized form
\begin{equation}
C^2+P^2+E^2=1.
\label{eq:triality}
\end{equation}
The variables $(C,P,E)$ therefore define coordinates on a
triality manifold, where $C$ and $P$ describe the cylindrical
projection of the reduced state and $E$ measures the radial deficit
arising from reduction of a larger Hilbert space.  In the pure-state
limit, $E=0$ and the dynamics remain on the Bloch-sphere surface.
Finite $E$ geometrizes mixedness as an emergent contraction
coordinate.
\paragraph{Dynamical constraint.}
Writing
\begin{equation}
x=C\cos\phi,
\qquad
y=C\sin\phi,
\qquad
z=P,
\label{eq:cylindrical}
\end{equation}
Eq.~\eqref{eq:bloch-eom} gives
\begin{align}
\dot C
&=
-\Gamma_2 C-gP\sin\phi,
\label{eq:Cdot}
\\
\dot\phi
&=
-\omega-\frac{gP}{C}\cos\phi,
\label{eq:phidot}
\\
\dot P
&=
gC\sin\phi-\Gamma_1(P-z_0).
\label{eq:Pdot}
\end{align}
Since $E^2=1-C^2-P^2$,
\begin{equation}
E\dot E
=
-C\dot C-P\dot P
=
\Gamma_2C^2+\Gamma_1P(P-z_0).
\label{eq:Edot}
\end{equation}
Hamiltonian precession therefore redistributes coherence and predictability but does not directly change the radial deficit. At steady state,
\begin{equation}
\Gamma_2C^2+\Gamma_1P(P-z_0)=0,
\label{eq:ss-constraint}
\end{equation}
or
\begin{equation}
C^2
=
-\frac{\Gamma_1}{\Gamma_2}P(P-z_0).
\label{eq:ss-C-P}
\end{equation}
The dynamics, therefore, do not explore the Bloch sphere uniformly,
but selects a lower-dimensional steady-state submanifold determined
by the competition between coherent precession and dissipation.

\paragraph{Work as holonomy.}
The geometric significance of the triality manifold becomes apparent
under quasistatic transport.
For a quasistatic protocol $\lambda(t)$, the work increment is
\begin{equation}
\delta W
=
\Tr\left[
\rhoSS(\lambda)\dd H(\lambda)
\right]
=
A_i(\lambda)\dd \lambda^i,
\label{eq:work-one-form}
\end{equation}
where
\begin{equation}
A_i(\lambda)
=
\Tr\left[
\rhoSS(\lambda)\partial_iH(\lambda)
\right]
\label{eq:connection}
\end{equation}
defines the work connection. For the Hamiltonian in Eq.~\eqref{eq:H},
\begin{equation}
A_\omega=\frac{P}{2},
\qquad
A_g=\frac{x}{2}
=
\frac{C\cos\phi}{2}.
\label{eq:A-components}
\end{equation}
Thus, for a closed quasistatic cycle $\mathcal C$,
\begin{equation}
W_{\rm cyc}
=
\oint_{\mathcal C}A
=
\frac{1}{2}
\oint_{\mathcal C}
\left(
P\,\dd\omega+C\cos\phi\,\dd g
\right).
\label{eq:cycle-work}
\end{equation}
By Stokes' theorem,
\begin{equation}
W_{\rm cyc}
=
\iint_{\Sigma(\mathcal C)}
F_{\omega g}\,\dd\omega\,\dd g,
\label{eq:stokes}
\end{equation}
with
\begin{equation}
F_{\omega g}
=
\partial_\omega A_g-\partial_gA_\omega
=
\frac{1}{2}
\left[
\partial_\omega(C\cos\phi)-\partial_gP
\right].
\label{eq:curvature-triality}
\end{equation}

If the dissipator is slaved to the instantaneous Hamiltonian eigenbasis, then
\begin{equation}
\rhoSS(\omega,g)=\rho_\beta[H(\omega,g)].
\end{equation}
In that case,
\begin{equation}
A_i
=
\Tr\left[
\rho_\beta\partial_iH
\right]
=
\partial_i F_{\rm th},
\label{eq:exact-gibbs}
\end{equation}
and therefore
\begin{equation}
F_{\omega g}=0,
\qquad
W_{\rm cyc}=0.
\label{eq:flat}
\end{equation}
Finite cyclic work, therefore, measures the failure of the
steady-state manifold to be globally integrable.
By contrast, when the dissipator is defined in a fixed pointer basis, $\rhoSS$ is generally not a Gibbs state of $H(\omega,g)$. The connection is then non-exact, the curvature is nonzero, and cyclic quasistatic protocols generate finite holonomic work.

\subsection{Phase-resolved triality curvature.}
For the fixed pointer-basis steady state,
\begin{equation}
\tan\phi=-\frac{\Gamma_2}{\omega}.
\label{eq:phase-relation}
\end{equation}
Together with Eq.~\eqref{eq:ss-C-P}, this allows the work curvature to be written in terms of the triality variables and the coherence phase:
\begin{equation}
F_{\omega g}
=
-\frac{
C\sin\phi
}{
2P\Gamma_1\Gamma_2 z_0
}
\left[
\Gamma_2 C^2
+
2\Gamma_2 P^2
-
\Gamma_1P^2\cos(2\phi)
\right].
\label{eq:F-CPE-phi}
\end{equation}
Using $C^2+P^2+E^2=1$, this becomes
\begin{equation}
F_{\omega g}
=
-\frac{
C\sin\phi
}{
2P\Gamma_1\Gamma_2 z_0
}
\left[
\Gamma_2(1-E^2+P^2)
-
\Gamma_1P^2\cos(2\phi)
\right].
\label{eq:F-CPE}
\end{equation}
The curvature is therefore not a scalar function of $(C,P,E)$ alone. The coherence phase $\phi$ is essential because holonomy is not
determined solely by the local triality coordinates, but by their
oriented transport within the pointer-basis plane. The triality relation constrains the local Bloch-sphere coordinates, while the phase-resolved curvature determines the finite cyclic work.

To remove the kinematic pole at $P=0$ and emphasize the
underlying geometric structure, we define the regularized
curvature
\begin{equation}
\widetilde F_{\omega g}
\equiv
P F_{\omega g}.
\label{eq:regularized-curvature}
\end{equation}
This regularization suppresses the coordinate singularity
associated with the cylindrical representation while
preserving the topology and localization structure of the
curvature field on the triality manifold.

Figure~\ref{fig:triality_curvature} visualizes the curvature field
associated with quasistatic work on the triality manifold.  The
regularized curvature localizes into symmetry-related interior sectors
rather than near extremal boundaries such as
$C=0$, $P=0$, or $E=0$.  Holonomic response is therefore maximized for
partially contracted states in which coherence, predictability, and
mixing coexist simultaneously.

Panel~(a) shows the regularized curvature field
$P F_{\omega g}$ painted directly onto the physical 
quarter-sphere
$C\ge0$ for maximal basis mismatch $\phi=\pi/2$.  The curvature
organizes into two symmetry-related response sectors separated by a
low-curvature region near the pure-state boundary $E=0$.  In this
sense, openness does not merely degrade geometric response; rather, the contraction induced by the environment becomes an essential
component of the geometry itself.  The localization of the curvature
away from extremal limits demonstrates that finite holonomic work
emerges from the combined transport of coherence, predictability, and
mixing over the triality manifold.

Panel~(b) shows a Mercator-like projection of the same curvature field
onto the coordinate chart $(\eta,\lambda)$ of the triality manifold.
Here, $\eta$ parameterizes azimuthal motion around the
coherence--predictability circle, while
\begin{equation}
E=\sin\lambda
\end{equation}
defines the contraction coordinate.  The dashed horizontal line marks
the equator $E=0$, corresponding to pure states without Bloch-sphere
contraction, while the vertical dashed line denotes the meridian
$\eta=0$ separating the two symmetry-related curvature sectors.  The
contour topology reveals two localized curvature basins connected by a
low-curvature corridor near the pure-state boundary, demonstrating
that the triality manifold possesses preferred geometric sectors in
which quasistatic work accumulates most efficiently.

Taken together, these figures show that complementarity in open quantum systems is not merely a kinematic constraint relation. Rather, the triality manifold supports an intrinsic curvature field whose geometry governs the holonomic response generated by quasistatic transport. Finite cyclic work therefore emerges directly from the holonomic transport of coherence, predictability, and openness

\subsection{Weak-Mismatch Expansion}
In the aligned limit $(\phi=0)$, the dissipative and Hamiltonian eigenbases coincide, the quasistatic work one-form becomes locally exact, and the curvature vanishes:
$ F_{\omega g}(\phi=0)=0.$
Nontrivial geometric response therefore emerges only when the pointer basis selected by the environment becomes misaligned with the coherent system dynamics. To clarify the physical origin of this effect, it is useful to examine the weak-mismatch regime $\phi\ll1$, where the curvature admits a regular perturbative expansion in the mismatch angle. Expanding the exact expression for $F_{\omega g}$ yields
\[
F_{\omega g}
\approx
\phi\left[
\Gamma_2(C^2+2P^2)-\Gamma_1P^2
\right]
+O(\phi^3),
\]
showing that the onset of geometric response is linear in the pointer–Hamiltonian mismatch. The curvature therefore appears continuously as the system departs from the locally exact aligned limit and does not rely on the extremal $\phi=\pi/2$ regime used in Fig.~\ref{fig:triality_curvature} 
to visualize the structure most clearly.

The structure of the linear coefficient is physically revealing. The term proportional to $\Gamma_2(C^2+2P^2)$ resembles the familiar transverse decoherence contribution appearing in the Bloch equations,
projected here onto the complementarity coordinates of the triality manifold.  In contrast, the $-\Gamma_1P^2$ contribution selectively suppresses curvature along the predictability direction, reflecting the competing influence of pure dephasing. Near the aligned limit, the geometric response is therefore governed by a balance between coherence-preserving and coherence-destroying dissipative channels. This establishes that the triality curvature is not merely an abstract geometric quantity, but rather directly reflects the same relaxation hierarchy that governs dissipative two-level dynamics in conventional open-system theory.

The weak-mismatch expansion also clarifies the origin of positive- and negative-curvature sectors observed in the full numerical results. Near $\phi=0$, the sign of the curvature is controlled by the competition between the coherence-associated contribution $\Gamma_2 C^2$ and the predictability contribution $(2\Gamma_2-\Gamma_1)P^2$. When $\Gamma_1<2\Gamma_2$, both sectors contribute with the same sign and the curvature remains predominantly positive. In contrast, for sufficiently strong pure dephasing $(\Gamma_1>2\Gamma_2)$, the predictability-dominated regions acquire opposite orientation, producing nodal lines and sign-changing curvature on the triality manifold. The appearance of positive and negative geometric sectors follows directly from the hierarchy of dissipative rates rather than solely from the global geometry of the manifold.

\section{Discussion}

The results presented here suggest that complementarity in open quantum systems is more than a static constraint on quantum states. Rather, it is the local coordinate expression of a geometric structure whose curvature governs physical response. In this framework, coherence, predictability, and entanglement are not independent resources but coordinates on a constrained manifold, while quasistatic work probes the holonomic transport of these variables over control space.

This perspective unifies two concepts that are typically treated
separately.  Complementarity is usually regarded as an
information-theoretic statement about what can be known or observed,
whereas work is a thermodynamic quantity associated with energy
exchange.  Here these notions are linked directly: the same variables that define the information budget of the system also determine the work connection, and their holonomic transport generates measurable cyclic work. In this sense, information constraints acquire a thermodynamic signature.

A central implication is the existence of a structural selection rule
for geometric work.  When the dissipative dynamics is slaved to the
instantaneous Hamiltonian eigenbasis, the steady state is Gibbsian, the
work connection is exact, and no cyclic quasistatic work can be
generated.  By contrast, when the dissipative pointer basis is fixed and misaligned with the Hamiltonian, the steady-state manifold develops curvature and supports holonomic transport. Finite cyclic work is therefore
not generic, but arises from a precise form of dynamical frustration
between coherent evolution and dissipation.  This identifies a class of
open quantum systems in which geometric work is symmetry-forbidden, and
a complementary class in which it is symmetry-allowed.

The geometric structure uncovered here also provides a new way to
interpret complementarity itself.  The triality relation
$C^2+P^2+E^2=1$ constrains the local coordinates of the state, but does
not determine how those coordinates evolve under control protocols.  The
curvature of the work connection encodes precisely this missing
information: it measures the failure of the complementarity variables to
be globally integrable.  Complementarity is therefore a local statement,
while holonomy captures its global consequence.

The geometric structure identified here suggests a broader reinterpretation of complementarity in nonequilibrium quantum physics. Traditionally, complementarity relations are viewed as local information-theoretic constraints imposed on quantum states. The present results show that, in open systems, these same constraints naturally define a curved manifold supporting holonomic transport. Observable quantities such as cyclic work therefore probe not only energetic response, but the global geometry of quantum information itself.

Perhaps most importantly, this framework makes complementarity operational. Rather than reconstructing density matrices or directly measuring coherence and entanglement, one can probe the curvature of the steady-state manifold through cyclic work measurements. The work performed over a closed loop provides a direct measure of the underlying geometric structure, and thus of the holonomic transport of the triality variables. This suggests a route toward ``cycle-based'' spectroscopy of open quantum systems, in which geometric response functions reveal information-theoretic structure.

More broadly, these results point to a geometric organization of
nonequilibrium quantum dynamics in which information constraints,
dissipation, and thermodynamic response are unified within a single
framework.  While we have demonstrated this explicitly for a driven
two-level system, the underlying structure is not model-specific.  In
higher-dimensional and many-body systems, similar geometric constraints
are expected to govern the interplay between coherence, entanglement,
and dissipation, with curvature providing a natural measure of
nonequilibrium structure.  
%In this sense, complementarity may be viewed
%as the local shadow of a more general geometric principle governing open
%quantum systems.
In this view, complementarity is not merely a limitation on
quantum information, but the local shadow of a geometric
principle governing transport, dissipation, and work in
open quantum systems.

%---connection to Bhavay's work
An especially interesting extension of the present framework concerns periodically driven and Floquet open quantum systems. In these systems, coherent evolution is generated not by a static Hamiltonian, but by time-periodic sequences of noncommuting operations whose stroboscopic dynamics can produce effective gauge structure, synchronization, and anomalous transport. The complementarity variables therefore evolve on a dynamically generated steady-state manifold whose geometry is determined jointly by Floquet driving and dissipation. Periodic driving can consequently generate effective curvature sectors even when the instantaneous dynamics remain locally integrable. Holonomic work and geometric response then become intrinsic properties of the Floquet transport itself, suggesting the possibility of engineering complementarity curvature through kicked or modulated quantum dynamics. More broadly, these considerations point toward a deeper connection between nonequilibrium geometry, Floquet transport, and observable-dependent response in driven open quantum systems.

\vspace{0.5cm}

\section*{Data Availability Statement}
All data generated or analyzed during this study are included in this manuscript.

\begin{acknowledgments}
This work at the University of Houston was supported by the National Science Foundation under CHE-2404788 and the Robert A. Welch Foundation (E-1337).
\end{acknowledgments}

\section*{Author Contributions}
The author developed the theoretical framework, performed all derivations, and carried out the analysis presented in this work.   The author thanks Bhavay Tyagi for useful discussions concerning periodically driven spin systems.

\section*{Conflicts of Interest}
The author declares no competing financial or non-financial interests.

\bibliographystyle{apsrev4-2}
\bibliography{geom_oqs_lit_v3,quantum_triality_master}

\end{document}